\documentclass[11pt]{article}
\usepackage{amsfonts}
\usepackage{amssymb}
\usepackage{amscd}
\usepackage{amsfonts}
\usepackage{amssymb}
\usepackage{amscd}

\def\bi{\begin{itemize}}
\def\ei{\end{itemize}}
\def\al{\alpha}
\def\bet{\beta}

\def\gam{\gamma }
\def\del{\delta}

\def\R{\mathbb{R}}
\def\dst{\displaystyle}
\def\vp{\vspace}
\def\hp{\hspace}

\def\noi{\noindent}
\def\nn{\nonumber}
\def\tr{\triangle}
\def\bea{\begin{eqnarray}}
\def\eea{\end{eqnarray}}
\def\ba{\begin{array}}
\def\ea{\end{array}}
\def\mhsd{\hspace{-2mm}}

\def\jzg{\mhsd = \mhsd}

\begin{document}

%\begin{center}
%\centerline
\title{\large\bf Path Integrals for Quadratic Lagrangians \\ on
$p$-Adic and Adelic Spaces}
\author{ Branko Dragovich\thanks{ Email address: \texttt{dragovich@ipb.ac.rs}}
 \\ \it Institute of Physics, Pregrevica 118, 11080 Zemun, Belgrade, Serbia
\\[7pt]  Zoran Raki\' c \thanks{ Email address: \texttt{zrakic@matf.bg.ac.rs}}
 \\ \it Faculty of Mathematics, University of Belgrade\\ \it Studentski trg 16, 11001 Belgrade, Serbia}
\date{}
%\end{center}

\maketitle

\begin{abstract}
Feynman's path integrals in ordinary, $p$-adic and adelic quantum
mechanics are considered. The corresponding probability amplitudes
${\cal K}(x^{''},t^{''};x',t')$ for two-dimensional systems with
quadratic Lagrangians are evaluated analytically  and obtained
expressions are generalized to any finite-dimensional spaces.
These  general formulas are presented in the form which is
invariant under interchange of the number fields ${\mathbb R}
\leftrightarrow {\mathbb Q}_p$ and ${\mathbb Q}_p \leftrightarrow
{\mathbb Q}_{p'} \, ,\, p\neq p'$. According to this invariance we
have that adelic path integral is a fundamental object in
mathematical physics of quantum phenomena.
\end{abstract}

\section{Introduction}

 To describe dynamics of a particle in classical mechanics, there
are Hamiltonian and Lagrangian formalisms which are equivalent.
Quantum mechanics is usually related to quantization of a
classical Hamiltonian consisting of a particle in an effective
field given by a potential.

Starting from the Hamiltonian there are two ways to treat quantum
evolution of a physical system: ({\it i}) the Heisenberg picture,
where time dependence is directly related  to the operator of an
observable $A$, i.e.
\bea i\hbar \frac{d\hat{A}}{dt} = i\hbar
\frac{\partial\hat{A}}{\partial t} + [\hat{A},\hat{H}],
\label{1.1} \eea
and ({\it ii}) the Schr\"odinger picture, where
time evolution is governed by the Schr\"odinger equation \bea
i\hbar\frac{\partial \Psi (x,t)}{\partial t} = H(\hat{k},x )\Psi
(x,t), \, \, \, \, \, \, \hat{k}= -i\hbar \frac{\partial}{\partial
x}. \label{1.2} \eea Both approaches are invented in the 1925-26
and shown to be equivalent versions of the same theory called
Quantum Mechanics.

Quantum mechanics related to the Lagrangian formalism started in
the 1932 by Dirac's observation that the quantum state in a point
$q+dq$ at the time $t+dt$ is connected with the state in the point
$q$ at $t$ by the transformation function $ \exp \left( \frac{ i\,
L\, dt}{\hbar}\right)\,,$ where $L=L(\dot{q},q,t)$ is the
classical Lagrangian. In the 1940's, Feynman developed Dirac's
approach and shown that dynamical evolution of the wave function $
\Psi (x,t) $ is \bea
   \Psi (x'',t'') = \int {\cal K}(x'',t'';x',t')\Psi(x',t')dx' ,
   \label{1.3}
\eea where \bea
   {\cal K}(x'',t'';x',t') = \int_{x',t'}^{x'',t''}
   \exp{\left(\frac{2\pi i}{h} \int_{t'}^{t''}
   L(\dot{q},q,t)dt\right)} {\cal D}q ,  \label{1.4}
\eea and $ \int_{t'}^{t''}
   L(\dot{q},q,t)\, dt = S[q]$ is the action for a path $q(t)$ connecting points $x'$
and $x''$. The integral in (\ref{1.4}) is known as the Feynman
path integral. In the Feynman definition \cite{feynm1},
discretizing the time $t$ into equidistant subintervals, the path
integral (\ref{1.4}) is the limit of the corresponding multiple
integral of $N$ variables $q_i = q(t_i), \ \ (i=1,2,...,N),$ when
$N\to\infty$. It is the primary object of the Feynman's path
integral approach to quantum mechanics which is related to the
classical Lagrangian formalism. Feynman's, Schr\"odinger's and
Heisenberg's approaches to ordinary quantum mechanics are
equivalent, but their formalisms are not equally suitable in some
generalizations.

 ${\cal K}(x'',t'';x',t')$ is the kernel of the corresponding
unitary integral operator $U(t'',t')$ acting as follows: \bea
   \Psi (t'') =  U(t'',t')\Psi (t') .              \label{1.5}
\eea $ {\cal K}(x'',t'';x',t')$ is also called  the probability
amplitude  for a quantum particle to pass from a  point $x'$ at
the time $t'$ to the other point $x''$ at $t''$. It is closely
related to Green's function and the quantum-mechanical propagator.

 Starting from (\ref{1.3}) one can easily derive the following
three general properties: \bea &&  \int{\cal K}(x'',t'';x,t) {\cal
K}(x,t;x',t') dx =
{\cal K}(x'',t'';x',t') ,      \label{1.6} \\[2pt]
&&
   \int \bar{{\cal K}}(x'',t'';x',t') {\cal K}(y,t'';x',t') dx' =
\delta (x''-y) ,       \label{1.7} \\[2pt]
&&
   {\cal K}(x'',t'';x',t'') = \lim_{t'\to t''} {\cal K}(x'',t'';x',t') =
\delta (x''-x') ,    \label{1.8} \eea  where integration is over
all the configuration space.

  For all its history, the path integral has been a subject of
great interest in theoretical and mathematical physics. It has
became, not only in quantum mechanics (see, e.g. \cite{stein1})
but also in the entire quantum theory, one of its the most
profound and suitable approaches to foundations and elaborations.
Feynman's path integral construction is also a natural and very
successful instrument in formulation and investigation of $p$-adic
\cite{volov1} and adelic \cite{brank1,branko1} quantum mechanics.
Moreover there are no $p$-adic analogs of the differential
equations (\ref{1.1}) and (\ref{1.2}).

 Adelic quantum mechanics contains complex-valued functions of
real and all $p$-adic arguments in the adelic form. There is not
the corresponding Schr\"odinger equation for $p$-adic dynamics,
but Feynman's path integral method is quite appropriate. Feynman's
path integral for probability amplitude in $p$-adic  quantum
mechanics ${\cal K}_p(x'',t'';x',t')$ \cite{volov1}, where ${\cal
K}_p$ is complex-valued and $x'',x',t'',t'$ are $p$-adic
variables, is a direct $p$-adic generalization of (\ref{1.4}),
i.e. \bea
   {\cal K}_p (x'',t'';x',t') = \int_{x',t'}^{x'',t''}
   \chi_p \left( -\frac{1}{h}
   \int_{t'}^{t''} L(\dot{q},q,t)dt\right) {\cal D}q ,
   \label{1.9}
\eea where $ \chi_p(a) = \exp{2\pi i\{a \}_p}$ is $p$-adic
additive character. The Planck constant $h$ in (\ref{1.4}) and
(\ref{1.9}) is the same rational number. We consider $p$-adic
valued integral $ \int_{t'}^{t''}L(\dot{q},q,t)dt$ as the
difference of antiderivative (without pseudoconstants) of
$L(\dot{q},q,t) $ in final $(t'')$ and initial $(t')$ times. In
the case of time discretization  we have ${\cal D}q =
\prod_{i=1}^N dq(t_i)$, where $dq(t_i)$ is the $p$-adic additive
Haar measure. Thus, $p$-adic path integral is the limit of the
multiple Haar integral when $N\to\infty$. To calculate (\ref{1.9})
in this way one has to introduce some ordering in the time $t\in
{\mathbb Q}_p$, and it is successfully done in  \cite{zelen1}. On
previous investigations of $p$-adic path integrals one can see
\cite{brank2,brank3,parisi,varadarajan} and references therein.
Some mathematical aspects of Feynman's path integral on real space
are vastly considered, e.g. see \cite{grosj1,grosj2,albev1}. Path
integral on $p$-adic space with $p$-adic valued probability
amplitude was considered in \cite{khr4}.

 Our main task here is an analytic evaluation of the $p$-adic
(\ref{1.9}) and the corresponding adelic Feynman path integrals
for the general case of Lagrangians $L(\dot{q},q,t)$, which are
quadratic polynomials in $\dot{q}$ and $q$, without making time
discretization. In fact, we will use the general requirements (in
particular, (\ref{1.6}) and (\ref{1.7})) which any (ordinary,
$p$-adic and adelic) path integral must satisfy. In some parts of
this evaluation, there is some similarity with Ref. \cite{grosj1}.
Adelic path integral may be viewed as an infinite product of the
ordinary one and $p$-adic path integrals for all primes $p$.
Formal definition of adelic path integral, with some of its basic
properties, will be presented in Section 5.

 Some of the main motivations to apply $p$-adic numbers and
adeles in quantum physics are: ({\it i}) the field of rational
numbers ${\mathbb Q}$, which contains all observational and
experimental numerical data, is the dense subfield not only in the
field of real numbers ${\mathbb R}$ but also in the fields of
$p$-adic numbers ${\mathbb  Q}_p$, ({\it ii}) there is  well
developed analysis (e.g. see \cite{volov2}) with $p$-adic valued
and complex-valued functions over ${\mathbb Q}_p$ which is
suitable in modern mathematical physics, ({\it iii}) general
mathematical methods and fundamental physical laws should be
invariant\cite{volov4} under an interchange of the number fields
${\mathbb R}$ and ${\mathbb Q}_p$, ({\it iv}) there is a quantum
gravity uncertainty $\Delta x$ of distances around the Planck
length $\ell_0$,
$
 \Delta x \geq \ell_0 = \sqrt{\hbar G/{c^3}} \sim 10^{-33} \mbox{cm} \,,
$ which restricts priority of archimedean geometry based on real
numbers and gives rise to employment of nonarchimedean geometry
related to $p$-adic numbers \cite{volov4}, and ({\it v}) it seems
to be quite natural to extend path integral on  real spaces to
adelic one by adding probability amplitudes over the paths on  all
$p$-adic spaces.

 Since 1987, there have been many publications (for a review,
see, e.g. \cite{volov2,freun1,khren1,DKKV} ) on possible
applications of $p$-adic numbers and adeles in  modern theoretical
and mathematical physics. The first successful employment of
$p$-adic numbers was in string theory. In Volovich's article
\cite{volov3}, a hypothesis on the existence of nonarchimedean
geometry at the Planck scale was proposed and $p$-adic string
theory was initiated.  Using $p$-adic Veneziano amplitude as the
Gel'fand-Graev \cite{gelfa1} beta function, Freund and Witten
obtained \cite{freun2} an attractive adelic formula, which states
that the product of the standard crossing symmetric Veneziano
amplitude and all its $p$-adic counterparts equals a constant.
Such approach gives a possibility to consider some ordinary string
amplitudes as an infinite product of their inverse $p$-adic
analogs. Many aspects of $p$-adic string theory have been of the
significant interest.

 For a systematic investigation of $p$-adic quantum dynamics, two
kinds of $p$-adic quantum mechanics have been formulated: with
complex-valued and $p$-adic valued wave functions of $p$-adic
variables (for a review, see \cite{volov1,volov2} and
\cite{khren1}, respectively). This paper is related to the first
kind of quantum mechanics, which can be presented as a triple \bea
    (L_2 ({\mathbb Q}_p), W, U(t)),
    \label{1.10}
\eea where $L_2({\mathbb Q}_p)$ is the Hilbert space on ${\mathbb
Q}_p$. $W$ denotes the Weyl quantization procedure and $U(t)$ is
the unitary representation of an evolution operator on
$L_2({\mathbb Q}_p)$. In our approach, $U(t)$ is naturally
realized by the Feynman path integral method. In order to connect
$p$-adic with standard quantum mechanics, adelic quantum mechanics
was formulated \cite{brank1}. Within adelic quantum mechanics a
few basic physical systems \cite{dragov1,dragov2}, including some
minisuperspace cosmological models \cite{cosmo1}, have been
successfully considered. As a result of $p$-adic effects in the
adelic approach, a space-time discreteness at the Planck scale is
obtained. Adelic path integral plays a central role and provides
an extension of contributions from quantum trajectories over real
space to probability amplitudes over paths in all $p$-adic spaces.
There have been also investigations on application of $p$-adic
numbers in the  spin glasses, Brownian motion, stochastic
processes, information systems, hierarchy structures, genetic
code, dynamics of proteins and some other phenomena related to
very complex dynamical systems (for a review see
\cite{volov2,khren1,paris1,khren2,DKKV}).

\section{ $p$-Adic  Numbers,  Adeles and Their Functions}

 In this section we give a brief review of some basic properties
of $p$-adic numbers, adeles, and their functions, which provides a
minimum of  mathematical background for next sections.

 There are physical and mathematical reasons to start with the
field of rational numbers ${\mathbb Q}$. From physical point of
view, numerical results of all experiments and observations are
some rational numbers,  i.e. they belong to ${\mathbb Q}$. From
algebraic point of view, ${\mathbb Q}$ is the simplest number
field of characteristic $0$. Recall that any $0\neq x\in {\mathbb
Q}$ can be presented as infinite expansions into the two different
forms: \bea
 x = \sum_{k=n}^{-\infty} a_k 10^k, \ \  a_k = 0,1,\cdots ,9,
\ \ a_n \neq 0,   \label{2.1} \eea which is the ordinary one to the
base $10$, and the other one to the base $p$ ($p$ is any prime
number) \bea
  x =  \sum_{k=m}^{+\infty} b_k p^k, \ \  b_k = 0,1,\cdots ,p-1,
\ \  b_m \neq 0,  \label{2.2}\eea where $n$ and   $m$ are some
integers which depend on $x$. The above representations
(\ref{2.1}) and (\ref{2.2}) exhibit the usual repetition of
digits, however the expansions are in the mutually opposite
directions. The series (\ref{2.1}) and (\ref{2.2}) are convergent
with respect to the metrics induced by the usual absolute value $|
\cdot |_\infty$ and $p$-adic norm $| \cdot |_p$, respectively.
Note that these valuations exhaust all possible inequivalent
non-trivial norms on ${\mathbb Q}$. Performing completions, i.e.
allowing all possible realizations of digits, one obtains standard
representation of real and $p$-adic numbers in the form
(\ref{2.1}) and (\ref{2.2}), respectively. Thus, the field of real
numbers ${\mathbb R}$ and the fields of $p$-adic numbers ${\mathbb
Q}_p$ exhaust all number fields which may be obtained by
completion of ${\mathbb Q}$, and which contain ${\mathbb Q}$ as a
dense subfield. Since $p$-adic norm of any term in (\ref{2.2}) is
$| b_k p^k |_p = p^{-k}$ if $b_k \neq 0$, geometry of $p$-adic
numbers is the nonarchimedean one, {\it i.e.} strong triangle
inequality $| x+y |_p \leq \mbox{max} (| x |_p, | y |_p) $ holds
and $| x |_p =p^{-m}$. ${\mathbb R}$ and ${\mathbb Q}_p$ have many
distinct algebraic and geometric properties.

 There is no natural ordering on ${\mathbb Q}_p$. However one can
introduce a linear order on ${\mathbb Q}_p$ in the following way:
$x < y$ if $| x |_p < | y|_p$, or if $| x|_p = | y|_p$ then there
exists such index $r \geq 0$ that digits satisfy $x_m = y_m,
x_{m+1} = y_{m+1}, \cdots, x_{m+r-1} = y_{m+r-1}, x_{m+r} <
y_{m+r}$. Here, $x_k$ and $y_k$ are digits related to $x$ and $y$
in expansion (\ref{2.2}). This ordering is very useful in time
discretization and calculation of $p$-adic functional integral as
a limit of the $N$-multiple Haar integral when $N\to \infty$.

 There are mainly two kinds of analysis on ${\mathbb Q}_p$ which
are of interest for physics, and they are based on two different
mappings: ${\mathbb Q}_p \to {\mathbb Q}_p$ and ${\mathbb Q}_p\to
{\mathbb C}$, where ${\mathbb C}$ is the field of ordinary complex
numbers. We use both of these analyses, in classical and quantum
$p$-adic models, respectively.

 Elementary $p$-adic valued functions and their derivatives are
defined by the same series as in the real case, but the regions of
convergence of these series are determined by means of $p$-adic
norm. As a definite $p$-adic valued integral of an analytic
function $f(x) = f_0 + f_1 x + f_2 x^2 + \cdots$ we take
difference of the corresponding antiderivative in end points, i.e.
$$
 \int_a^b f(x) = \sum_{n=0}^\infty \frac{f_n}{n+1} \left( b^{n+1} -
 a^{n+1}   \right).
$$

 Usual complex-valued functions of $p$-adic variable, which are
employed in mathematical physics, are: ({\it i}) an additive
character $\chi_p (x) = \exp{2\pi i \{x \}_p}$, where $\{x \}_p$
is the fractional part of $x\in {\mathbb Q}_p$, ({\it ii}) a
multiplicative character $\pi_s (x) = |x|_p^s$, where $s\in
{\mathbb C}$, and ({\it iii}) locally constant functions with
compact support, like $\Omega (|x|_p)$, where \bea \Omega (|x|_p)
= \left\{
\begin{array}{ll}
                 1,   &   |x|_p \leq 1,  \\[3pt]
                 0,   &   |x|_p > 1.
                 \end{array}    \right.
                 \label{2.3}
\eea There is well defined Haar measure and integration. So, we have
\bea
 && \int_{{\mathbb Q}_p} \chi_p (ayx)\, dx = \delta_p (ay) = |a|^{-1}_p\,
  \delta_p (y) , \ \ a\neq 0,
  \label{2.4a} \\[2pt]
&&
  \int_{{\mathbb Q}_p} \chi_p (\alpha x^2 + \beta x) \, dx
  =\lambda_p (\alpha)\, |2\alpha|_p^{-\frac{1}{2}}
 \, \chi_p \left( -\frac{\beta^2}{4\alpha} \right), \ \ \alpha\neq
  0,
                                                   \label{2.4b}
\eea where $\delta_p (u)$ is the $p$-adic Dirac $\delta$-function.
The number-theoretic function $\lambda_p (x)$ in (\ref{2.4b}) is a
map $\lambda_p: {\mathbb Q}_p^\ast\to {\mathbb C}$ defined as
follows \cite{vladimirov}: \bea && \lambda_p (x) = \left\{
\begin{array}{lll}
                 1,    &  m=2j, & \quad  p\neq 2,  \\[3pt]
               \sqrt{\left(\frac{-1}{p} \right)}  \left(\frac{x_m}{p} \right),    &  m=2j+1, & \quad p \neq 2,
% \ \  p\equiv 1(\mbox{mod}\ 4), \\[3pt]
  %               i\left(\frac{x_m}{p} \right),    &  m=2j+1,
 % \ \  p\equiv 3(\mbox{mod}\ 4),
                 \end{array}   \right.            \label{2.5}\\[7pt]
                 &&
  \lambda_2 (x) = \left\{   \begin{array}{ll}
                  \exp{[\pi i (1/4 + x_{m+1})]},
   &  m= 2j,  \\[5pt]
                 \exp{[\pi i (1/4 + x_{m+1}/2 + x_{m+2})]},      &   m=2j+1,
                 \end{array}
                 \right.        \label{2.6}
\eea where $x$ is presented in the form (\ref{2.2}),  $j\in {\mathbb
Z}$,
 $\left(\frac{x_m}{p} \right)$ is the  Legendre symbol defined as
\bea
 \left(\frac{a}{p} \right) = \left\{   \begin{array}{ll}
                 1,    &  \mbox{if} \ \ \ a\equiv y^2(\mbox{mod}\ p)\, ,  \\
                -1,    &  \mbox{if} \ \ \ a\not\equiv y^2(\mbox{mod}\ p) \, ,
                 \end{array}   \right.            \label{2.7}
\eea and ${\mathbb Q}_p^\ast = {\mathbb Q}_p\setminus \{0\}$. We
will also take $\lambda_p (0) = 1$. It is often sufficient to use
standard properties:
$$
  \lambda_p (a^2 x)= \lambda_p (x), \quad \lambda_p (x) \lambda_p (-x) =1,
\quad   \lambda_p \left(\frac{xy}{x+y} \right) = \frac{\lambda_p
(x) \, \lambda_p (y)}{\lambda_p (x+y)},
$$
\bea  \lambda_p(x) \, \lambda_p(y) = (x,y)_p \, \lambda_p(xy)\,
\lambda_p(1),  \quad \quad |\lambda_p (x)|_\infty = 1, \qquad
a\neq 0, \label{2.8} \eea
where $(x,y)_p$ is the Hilbert symbol.
Recall that the Hilbert symbol $(a,b)_p \, , \quad a,b\in {\mathbb
Q}_p ,$ is $+ 1$ or $- 1$ if there exist such $x,y,z \in {\mathbb
Q}_p $ that equation $a\, x^2 + b\, y^2 = z^2$ has or has not a
nontrivial solution, respectively.

Recall that the real analogs of  (\ref{2.4a}) and (\ref{2.4b})
have the same form , {\it i.e.} \bea
 && \int_{{\mathbb Q}_\infty} \chi_\infty (ayx)\, dx = \delta_\infty (ay)
  = |a|^{-1}_\infty\,
  \delta_\infty (y) \,, \ \ a\neq 0,
  \label{2.9a} \\[3pt]
&&
  \int_{{\mathbb Q}_\infty} \chi_\infty (\alpha x^2 + \beta x)\, dx
  =\lambda_\infty (\alpha)\, |2\alpha|_\infty^{-\frac{1}{2}}
 \, \chi_\infty \left( -\frac{\beta^2}{4\alpha} \right)\,, \ \ \alpha\neq 0,
                                                   \label{2.9b}
\eea where ${\mathbb Q_\infty}\equiv {\mathbb R}$, $\, \,
\chi_\infty (x) = \exp{(-2\pi i x)}$ is additive character in the
real case and $\delta_\infty$ is the ordinary Dirac
$\delta$-function. Function $\lambda_\infty (x)$ is defined as
\bea
  \lambda_\infty (x) = \exp{\left[- \pi i \frac{\mbox{sgn}\ x} {4}\right]}, \ \
  x\in {\mathbb R}^\ast ={\mathbb R}\setminus \{ 0 \} \label{2.10}
\eea and exhibits the same properties (\ref{2.8}), i.e. equalities
(\ref{2.8}) have place if we replace index $p$ by $\infty$. In the
real case, the Hilbert symbol $(x,y)_\infty$ is equal to $-1$ if
$x< 0, \, y< 0$ and otherwise is $+1$.

 Since we are interested in Feynman's path integral on spaces
with any finite number of dimensions, generalization of some
previous formulas has to be introduced.

\bigskip

\noindent{\bf Definition 2.1.} Let \bea
 \Lambda_v (x_1\ , x_2\ , \cdots\ , x_n) = \prod_{i=1}^n \, \lambda_v
(x_i)
                                                   \label{2.11}
\eea be new number-theoretic functions, where subscript $v = \infty,
2, 3,\cdots, p, \cdots$ denotes real as well as any $p$-adic case.
\bigskip

\noindent{\bf Proposition 2.2.} {\it The new  functions $
\Lambda_v (x_1\ , x_2\ , \cdots \ , x_n)$ satisfy the following
property: \bea   \Lambda_v (x_1\,, x_2\,, \cdots\,, x_n) =
\lambda_v (x_1 x_2 \cdots x_n) \lambda_v^{n-1} (1) \prod_{i<j\leq
n} (x_i,x_j)_v  \,. \label{2.12} \eea }

\bigskip

\noi{\it Proof. \,\,} Formula (\ref{2.12}) follows from the above
property $\lambda_v (a)\, \lambda_v (b) = (a,b)_v \, \lambda_v (1)
\, \lambda_v (a b) $ and the properties of the Hilbert symbol:
$(a, b)_p=(b, a)_p$ and $(a, b c)_p=(a, b)_p \, (a, c)_p$, see
\cite{vladimirov}. $\qquad \Box$

\bigskip

\noindent{\bf Proposition 2.3.} {\it Let $x = (x_1\,,x_2 \,, \cdots,
x_n) , \, y = (y_1\,,y_2\,, \cdots, y_n) $ be column vectors, and
let $ B = (B_{kl})$ be a nonsingular $n\times n$ matrix, where
$x_k,\, y_k, \, B_{kl} \in {\mathbb Q}_v$. Then\bea \int_{{\mathbb
Q}^n_v} \chi_v (y^T B x) \, d^n x = |\det (B_{kl}) |_v^{-1}
\prod_{k=1}^n \delta_v (y_k) \,,          \label{2.13} \eea where
${y}^T$ denotes transpose map of $y$}.

\bigskip

\noi{\it Proof. \, \,}  Let us  change variables of integration by
$z_k =\sum_{l=1}^n B_{kl} x_l$. Then we have $d^nz =|\det
(B_{kl})|_v \, d^nx. $  The integral (\ref{2.13}) can be rewritten
as $|\det (B_{kl})|^{-1}_v \prod_{k=1}^n \int_{{\mathbb Q}_v}
\chi_v (y_kz_k) \, dz_k$. According to (\ref{2.4a}) and
(\ref{2.9a}), we obtain (\ref{2.13}).     $ \qquad \Box $

\bigskip

\noindent{\bf Proposition 2.4.}  {\it Let $x = (x_1,x_2, \cdots,
x_n), $ $ \, \beta = (\beta_1,\beta_2, \cdots, \beta_n)$ be two
column vectors, and let $ \alpha = (\alpha_{kl})$ be a nonsingular
symmetric $n\times n$ matrix, where $x_k,\, \beta_k, \, \alpha_{kl}
\in {\mathbb Q}_v$. Then \bea \int_{{\mathbb Q}^n_v} \chi_v (x^T
\alpha x + \beta^T x)\, d^n x = \Lambda_v (\alpha_1,\alpha_2,\cdots
,\alpha_n)\, |\det (2\alpha_{kl}) |_v^{-\frac{1}{2}} \, \chi_v
\left( -\frac{1}{4} \beta^T \alpha^{-1} \beta \right) , \label{2.14}
\eea where $\alpha_1, \alpha_2, \cdots, \alpha_n$ are eigenvalues of
the matrix $\alpha$}.

\bigskip

\noi{\it Proof. \,\,}  Consider first the case $\beta = 0$. Using
an orthogonal rotation $n\times n$ matrix $A$ such that $x' = A x$
and $x^T \alpha x = x'^T \alpha' x'$, where $\alpha' = A\alpha A^T
= diag (\alpha_1,\alpha_2, \cdots, \alpha_n) ,$ one obtains
\bea\nn
 \int_{{\mathbb Q}^n_v} \chi_v (x^T \alpha x )\, d^n x &\hp{-2mm}=\hp{-2mm}& \prod_{k=1}^n
 \int_{{\mathbb Q}_v} \chi_v (x'_k \alpha_k x'_k )\, dx'_k = \prod_{k=1}^n
  \lambda_v (\alpha_k)\,|2\alpha_k|_v^{-\frac12}\\ \nn &\hp{-2mm}=\hp{-2mm}&
 \Lambda(\alpha_1,\alpha_2,\dots,\alpha_n)|\, \det
(2\alpha_{kl}) |_v^{-\frac{1}{2}}\,. \eea Employing (\ref{2.4b}) and
(\ref{2.9b}), as well as (\ref{2.11}) and the property that the
determinant of a matrix is the product of all its eigenvalues, we
gain  (\ref{2.14}) for $\beta = 0$. The final result follows from
the identity
$$
x^T \alpha\, x + \beta^T x = (x + \frac{1}{2} \alpha^{-1} \beta)^T
\alpha \, (x + \frac{1}{2} \alpha^{-1} \beta ) -\frac{1}{4}
\beta^T \alpha^{-1} \beta
$$
and after shifting the integration variable.      $ \qquad \Box $

\bigskip

\noindent{\bf Remark 2.5.} Since the determinant of a matrix is
the product of all its eigenvalues, it is worth noting that
according to (\ref{2.12}) one can express $\Lambda_v
(\alpha_1\alpha_2\cdots \alpha_n) $ in (\ref{2.14}) in the
following form: \bea
 \Lambda_v (\alpha_1\,,\alpha_2\,,\cdots\,, \alpha_n) = \lambda_v (\det (\alpha_{kl}))
               \, \lambda_v^{n-1} (1) \, \prod_{i<j\leq n} (\alpha_i,\alpha_j)\,.
                           \label{2.15}
\eea

\bigskip

 For  more  information on  usual properties of $p$-adic numbers
and related analysis one can see \cite{volov2,gelfa1,schik1}.

 Real and $p$-adic numbers are unified in the form of adeles. An
adele $x$ \cite{gelfa1} is an infinite sequence \bea
  x= (x_\infty, x_2, \cdots, x_p, \cdots),             \label{2.16}
\eea where $x_\infty \in {\mathbb R}$ and $x_p \in {\mathbb Q}_p$
with the restriction that for all but a finite set $\mathcal{P}$
of primes $p$ one has  $x_p \in {\mathbb Z}_p$, where ${\mathbb
Z}_p = \{ a\in {\mathbb Q}_p: |a|_p \leq 1  \}$ is the ring of
$p$-adic integers. Componentwise addition and multiplication are
natural operations on the ring of adeles $\mathbb{A}$, which can
be regarded as \bea
 \mathbb{A} = \bigcup_{\mathcal{P}} \mathbb{A} (\mathcal{P}),
 \qquad \quad  \mathbb{A}(\mathcal{P}) = {\mathbb R}\times \prod_{p\in \mathcal{P}} {\mathbb Q}_p
 \times \prod_{p\not\in \mathcal{P}} {\mathbb Z}_p.         \label{2.17}
\eea      $\mathbb{A}$ is a locally compact topological space.

 There are also two kinds of analysis over topological ring of
adeles $\mathbb{A}$, which are generalizations of the
corresponding analyses over ${\mathbb R}$ and  ${\mathbb Q}_p$.
The first one is related to mapping $\mathbb{A} \to \mathbb{A}$
and the other one to $\mathbb{A} \to \mathbb{C}$. In
complex-valued adelic analysis it is worth mentioning an additive
character \bea
 \chi (x) = \chi_\infty (x_\infty) \prod_p \chi_p (x_p),
 \label{2.18}
\eea a multiplicative character \bea
  |x|^s = |x_\infty|_\infty^s \prod_p |x_p|_p^s, \ \ s\in \mathbb{C},
                                                    \label{2.19}
\eea and elementary functions of the form \bea
 \phi (x) = \phi_\infty (x_\infty) \prod_{p\in \mathcal{P}} \phi_p (x_p)
 \prod_{p\not\in \mathcal{P}} \Omega (|x_p|_p),           \label{2.20}
\eea where $\phi_\infty (x_\infty)$ is an infinitely
differentiable function on ${\mathbb R}$ such that $|x_\infty
|_\infty^n \phi_\infty (x_\infty) \to 0$ as $|x_\infty|_\infty \to
\infty$ for any $n\in \{0,1,2,\cdots  \}$, and $\phi_p (x_p)$ are
locally constant functions with compact support. All finite linear
combinations of elementary functions (\ref{2.20}) make the set
$S(\mathbb{A})$ of the Schwartz-Bruhat adelic functions. The
Fourier transform of $\phi (x)\in S(\mathbb{A})$, which maps
$S(\mathbb{A})$ onto $S(\mathbb{A})$, is \bea
 \tilde{\phi}(y) = \int_{\mathbb{A}} \phi (x)\chi (xy)dx,
 \label{2.21}
\eea where $\chi (xy)$ is defined by (\ref{2.18}) and $dx =
dx_\infty dx_2 dx_3 \cdots$ is the Haar measure on $\mathbb{A}$.

It is worth mentioning the following adelic products
\cite{vladimirov}: \bea
 &&\chi_\infty (x) \prod_p \chi_p (x) = 1,  \quad x \in \mathbb{Q} \\
&&|x|_\infty^s \prod_p |x|_p^s  = 1,  \quad x \in \mathbb{Q}^\ast,
 \, \, s\in \mathbb{C} \\
 && \lambda_\infty (x) \prod_p \lambda_p (x) = 1, \quad x \in \mathbb{Q}^\ast \\
 && (x, y)_\infty \prod_p (x, y)_p = 1,   \quad x,y \in \mathbb{Q}^\ast .
 \label{2.22}
\eea

 One can define the Hilbert space on $\mathbb{A}$, which we will
denote by $L_2(\mathbb{A})$. It contains infinitely many
complex-valued functions of adelic argument (for example,
$\Psi_1(x), \Psi_2(x), \cdots$) with scalar product
$
 (\Psi_1,\Psi_2) = \int_{\cal A} \bar{\Psi}_1(x) \Psi_2(x) dx
$ and norm $
  ||\Psi|| = (\Psi,\Psi)^{\frac{1}{2}} < \infty \,,
$
where $dx$ is the Haar measure on $\mathbb{A}$. A basis of
$L_2(\mathbb{A})$ may be given by the set of orthonormal
eigefunctions in spectral problem of the evolution operator $U
(t)$, where $t\in \mathbb{A}$. Such eigenfunctions have the form
\bea
 \psi_{\mathcal{P},\alpha} (x,t) = \psi_n^{(\infty)}(x_\infty,t_\infty)
 \prod_{p\in \mathcal{P}} \psi_{\alpha_p}^{(p)} (x_p,t_p)
 \prod_{p\not\in \mathcal{P}} \Omega (|x_p|_p),
 \label{2.22}
\eea where $\psi_n^{(\infty)}$ and $\psi_{\alpha_p}^{(p)}$ are
eigenfunctions in ordinary and $p$-adic cases, respectively.
$\Omega (|x_p|_p)$ is defined by (\ref{2.3}) and presents a state
invariant under transformation of $U_p(t_p)$ evolution operator.
Adelic quantum mechanics \cite{brank1,branko1} may be regarded as
a triple
$$
    (L_2 (\mathbb{A}), W(z), U(t)),
$$
where $W(z)$  and $U(t)$ are unitary representations of the
Heisenberg-Weyl group and evolution operator on $L_2
(\mathbb{A})$, respectively.

\section{Quadratic Lagrangians and Their Actions}

  A general quadratic Lagrangian
can be written in matrix form as follows: \bea
%  &&  L(\dot{q},q,t) = \frac{1}{2}\, \dot{q}_k \, \alpha_{kl} \, \dot{q}_l
%   +  \dot{q}_k \, \beta_{kl} \, {q}_l
%   + \frac{1}{2}\, {q}_k \, \gamma_{kl} \, {q}_l
%   + \delta_k \, \dot{q}_k
%   + \eta_k \, {q}_k
%   + \varepsilon \\
    \label{3.1} L(\dot{q},q,t) = \frac{1}{2}\, \dot{q}^T \, A \, \dot{q}
   +  \dot{q}^T \, B \, {q}
   + \frac{1}{2}\, {q}^T \, C \, {q}
   + D^T \, \dot{q}
   + E^T \, {q}
   + \varepsilon               %\eqno(3.1)
\eea  where $A=(\al_{kl} (t))$ is a regular symmetric matrix,
$C=(\gam_{kl} (t))$ is a symmetric matrix, $B=(\bet_{kl} (t))$ is
a matrix, $D=(\del_{k} (t))$, $E=(\eta_{k} (t))$, $q=(q_k(t))$ and
$\dot{q}=(\dot{q_k}(t))$ are vectors in $\R^n$.  All matrices are
of type $n\times n$ with matrix elements viewed as analytic
functions of the time $t$. In fact, we want to consider the
corresponding adelic Lagrangian, i.e. an adelic collection of
Lagrangians of the same form (\ref{3.1}) which differ only by
their valuations $v = \infty, 2, 3,\cdots$. In this section we
present some results valid simultaneously for real as well as for
$p$-adic classical mechanics. In adelic case their power series
expansions will have the same rational coefficients in the real
and all $p$-adic cases.

 The Euler-Lagrange equations of motion are \bea
%&&
%\alpha_{kl}\, {\ddot q}_l + [\dot{\alpha}_{kl}\, + (\beta_{kl} -
%\beta_{kl})] {\dot q}_l +
% ( \dot{\beta}_{kl}\,   - \gamma_{kl})\,  q_l  = \eta_k - \dot{\delta}_k ,
% \quad \quad k= 1,2,\cdots, n\,.
 % \\[2pt]
  && \label{3.2}
 A\, {\ddot q} + (\dot{A} + B - B^T)\, {\dot q} +
 ( \dot{B}\,   - C)\,  q  = E - \dot{D} .
\eea Generally, (\ref{3.2}) represents a system of $n$ coupled
linear inhomogeneous differential equations of the second order.
When it is coupled, starting from the homogeneous system and
eliminating derivatives of all but one coordinate, one can
construct a system of $n$ uncoupled (resolvent) homogeneous linear
differential equations of the $2n$ order. Thus a general solution
of (\ref{3.2}), which describes classical trajectory, can be found
by means of solution of the corresponding uncoupled equations. In
this way we have \bea && q_k = x_k(t) = \sum_{m=1}^{2n}
f_{km}(t)\, C_m + \xi_k (t) , \label{3.3}\quad \mbox{or}\quad q =
x(t) = F(t)\, \mathcal{C} + \xi (t), \eea where   $
F(t)=[f_{km}(t)]\in M_{n,2n}$ is a solution of the corresponding
system of homogeneous differential equations,
$\mathcal{C}=[C_m]\in M_{2n,1}$ is the vector of constants, and
$\xi (t)=[\xi_k(t)]\in M_{n,1}$  is a particular solution of the
complete system of differential equations (\ref{3.2}). If we
choose $f_{1m}(t)$ as linearly independent solutions for $x_1 (t)$
then solutions $f_{km}(t)$ for $x_k (t), \,$ $k\neq 1, \,$ are
determined by the system (\ref{3.2}) and they are related to
$f_{1m}(t)$.

 For the boundary conditions $x'_k=x_k(t')$ and $x''_k=x_k(t'')$,
let us introduce the following useful notations: \bea &&\hp{-15mm}
f_i(t)=[f_{i1}(t),\dots,f_{i\,2n}(t)],\quad
f^i(t'',t')=[f_{1i}(t''),\dots,f_{ni}(t''),f_{1i}(t')\dots,f_{n\,i}(t')]^T\,,\\[2pt]
&&\hp{-15mm} f''_i=f_i(t''),\quad f'_i=f_i(t'), \quad
\dot{f''}_i=\dot{f}_i(t''),\quad\dot{f'}_i=\dot{f}_i(t'), \\[3pt]
&& \hp{-15mm}\mathcal{F}=\mathcal{F}(t'',t')=\left[\ba{c} F(t'') \\
F(t')\ea\right]=\left[\ba{c} F'' \\
F'\ea\right]=\left\{\ba{l}
[f_1(t''),\dots,f_n(t''),f_1(t'),\dots,f_n(t')]^T \\[5pt] [f^1(t'',t'),f^2(t'',t'),\dots,f^{2n}(t'',t')],\ea\right.  \\
&& \nn \hp{-15mm} \mbox{where $[f_1(t''),..,f_n(t''),f_1(t'),..,f_n(t')]^T$ is a matrix with rows $f_1(t''),..,f_n(t')$} \\[4pt]
&& \hp{-15mm}\tr=\tr(t'',t')=\det\mathcal{F},\ \
\mathcal{F}_{ij}=\mbox{$(ij)$-algebraic complement of
$\mathcal{F}$},\ \ \tr_{i,j}=\det\mathcal{F}_{ij},
\\[5pt] &&\hp{-15mm} {F}''=F(t''),\quad {F}'=F(t'),\quad
x\xi=[x_1''-\xi_1'',\dots,x_n''-\xi_n'',x_1'-\xi_1',\dots,x_n'-\xi_n']^T
\\[3pt]
&&\hp{-15mm} \tr_i=\tr_i(t'',t')=\det[f^1(t'',t'),\dots,
f^{i-1}(t'',t'),x\xi,f^{i+1}(t'',t')\dots,f^{2n}(t'',t')]\,,\\[5pt] &&\hp{-15mm} \dot{\tr}_i(f'_j)(t'',t')= \dot{\tr}_i(f'_j)=
\det[\,f''_1,..,
f''_{i-1},\dot{f}'_j,f''_{i+1},..,f''_{n},f'_1,..,f'_n\,]\,,
i,j=1,..,n,\\[5pt] &&\hp{-15mm} \dot{\tr}_{i+n}(f''_j)(t'',t')=\dot{\tr}_{i+n}(f''_j)=
\det[\,f''_1,..,f''_n, f'_1,..,
f'_{i-1},\dot{f}''_j,f'_{i+1},..,f'_n\,], i,j=1,..,n.\eea

\bigskip

\noindent{\bf Proposition 3.1.} {\it Imposing the boundary
conditions $x'_k=x_k(t')$ and $x''_k=x_k(t'')$, vector of
constants of integration $\mathcal{C}$  becomes: \bea \mathcal{C}=
\mathcal{C} (t'',t') =
\frac{1}{\tr(t'',t')}\;[\tr_1(t'',t'),\tr_2(t'',t'),\dots,\tr_{2n}(t'',t')]^T
. \label{3.4} \eea \smallskip}

\bigskip

\noi{\it Proof. \, \,} It follows after performing relevant
computations.
      $ \qquad \Box $

\bigskip

 Note that in the real case for periodic solutions $ f_{km}(t+T)
= f_{km}(t)$, the determinant $\tr$ can be singular. To avoid such
problem one has then to restrict the time interval $t''-t'$ to be
smaller than the period $T$.

 Taking into account (\ref{3.4}), one can rewrite (\ref{3.3}) in
the following form \bea \nn && x_k(t) = \frac{1}{\tr(t'',t')}
\sum_{i=1}^{2n}\, \tr_i(t'',t'){f_{ki}(t)} + \xi_k (t)\,,\qquad
k=1,2,\dots,n\,. \eea

 Using the equations of motion (\ref{3.2}), the Lagrangian
(\ref{3.1}) can be rewritten as \bea
  L(\dot{x},x,t) = \frac{1}{2}\frac{d}{dt}
  \left[ x^T A\,\dot{x} + x^T\, B\,x + D^T x  \right] +
  \frac{1}{2} \left( D^T \dot{x} + E^T x  \right) +
  \varepsilon
                            \label{3.7}
\eea where $x(t)$ denotes now the classical trajectory (\ref{3.3}).

Using method which was described in \cite{grosj2} in the case
$n=3$ one can find the following very important result.

\bigskip

 \noindent{\bf Theorem 3.2.} Let $ \{f_{1j}\;,j=1,2,\dots,2\,n\}$ be any
 linearly independent solutions of the resolvent equation for
 $x_1(t)$ in (\ref{3.2}),  then solutions $f_{km}(t)$ for $x_k (t), \,$
$k\neq 1, \,$ are determined by the system (\ref{3.2})  and
the following equality holds \bea \label{sols} \det\left[\ba{c} F(t)\\
\dot{F}(t) \ea\right]=\frac{\mathcal{D}}{\det A},\eea where
$\mathcal{D}$ is a non-zero constant, which could be chosen to be
equal  $1$.

\bigskip

\noindent{\bf Proposition 3.3.} {\it The general form of the action
for classical trajectory $x(t)$ of a quadratic Lagrangian, for a
particle being in point $x'$ at the time $t'$ and in position $x''$
at $t''$, is \bea
 \bar{S}(x'',t'';x',t') = \frac{1}{2}\,x''^T\, \bar{A}\, x'' +  x''^T \bar{B}\, x' +
 \frac{1}{2}\,x'^T \bar{C}\, x' +  \bar{D}^T x''+  \bar{E}^T\, x'
 + \bar{\varepsilon},
                                                         \label{3.8}
\eea where $\bar{A}=[\bar{A}_{kl}], \,\bar{B}=[\bar{B}_{kl}],\,
\bar{C}=[\bar{C}_{kl}],\, \bar{D}=[\bar{D}_k] ,\,$ and $
\bar{E}=[\bar{E}_k] .$        \bea \nn && \hp{-8mm} \bar{A}_{kl} =
\bar{A}_{kl}(t'',t') = \frac{\partial^2 \bar{S}_0}{\partial x''_k
  \partial x''_l} , \quad
  \bar{B}_{kl} = \bar{B}_{kl}(t'',t') = \frac{\partial^2
\bar{S}_0}{\partial x''_k
  \partial x'_l} , \quad
  \bar{C}_{kl} = \bar{C}_{kl}(t'',t') = \frac{\partial^2
\bar{S}_0}{\partial x'_k
  \partial x'_l} , \\[5pt]
&&  \hp{-8mm}
 \bar{D}_k = \bar{D}_k (t'',t')= \frac{\partial\bar{S}_0}{\partial x''} ,
 \quad
\bar{E} = \bar{E}_k (t'',t')= \frac{\partial\bar{S}_0}{\partial
x'} , \quad \bar{\varepsilon}=  \bar{\varepsilon} (t'',t') =
\bar{S}_0 \nn \eea and subscript ${}_0$ in the classical action
means that after performing  derivatives of the
$\bar{S}(x'',t'';x',t')$ one has to replace $x''$ and  $x'$ by
$x'' = x' = 0$.}

\bigskip

\noi{\it Proof. \, \,} From (\ref{3.4}) it is clear that constants
of integration $C_i(t'',t')$ are linear in $x''_k$  and $x'_l$.
Then the corresponding classical action \bea \nn
\bar{S}(x'',t'';x',t') &\jzg& \int_{t'}^{t''} L(\dot{x},x,t)\, dt
=
  \frac{1}{2}
  \left[ x^T A\,\dot{x} + x^T B\, x + D^T\, x  \right]
  \,\mid_{t'}^{t''} \\[3pt]
&& +\
  \frac{1}{2}  \int_{t'}^{t''} \left( D^T  \dot{x}+ E^T x  \right)\,dt +
  \int_{t'}^{t''} \varepsilon (t) \,dt
  \label{3.10}\eea
 is quadratic in $x''_k$ and $x'_l$, where subscripts run again the
same values,  $k,\,l = 1,\cdots, n$. $ \qquad \Box $

\bigskip

 For our evaluation of the path integrals it is especially
important to have explicit dependence of coefficients
$\bar{A}_{kl},\, \bar{B}_{kl} $ and $ \bar{C}_{kl}$ on
coefficients in the Lagrangian (\ref{3.1}) and on ingredients of
the classical trajectory (\ref{3.3}). Because of that, we can
rewrite (\ref{3.10}) in the following way:  \bea \label{lin}
\bar{S}(x'',t'';x',t') &\jzg& \int_{t'}^{t''} L(\dot{x},x,t)\, dt
=
  \frac{1}{2}
  \left[\, x^T A\,\dot{x} + x^T B\, x \,\right]\mid_{t'}^{t''} +\ \mathcal{L}{in}(x'',x')
  \\[3pt] \nn &\jzg&\frac{1}{2}\,[\,(\mathcal{C}^TF(t'')^T +\xi(t'')^T) A(t'')\,(\mathcal{C}\dot{F}(t'')+\dot{\xi}(t''))
  + (\mathcal{C}^TF(t'')^T +\xi(t'')^T) \\[5pt] \nn &&\times\, B(t'')(\mathcal{C}F(t'')+{\xi}(t'')) - (\mathcal{C}^TF(t')^T +\xi(t')^T) A(t')
  \,(\mathcal{C}\dot{F}(t')+\dot{\xi}(t'))
   \\[5pt] \nn && +\ (\mathcal{C}^TF(t')^T+ \xi(t')^T) B(t')(\mathcal{C}F(t')+{\xi}(t')) ]+\
   \mathcal{L}{in}(x'',x')\\[5pt] &\jzg& \frac{1}{2}\,[\,\mathcal{C}^TF(t'')^T A(t'')\,\mathcal{C}\dot{F}(t'')
  + \mathcal{C}^TF(t'')^T B(t'')\mathcal{C}F(t'') \nn \\[5pt] \nn && -\ \mathcal{C}^TF(t')^T A(t')
  \mathcal{C}\dot{F}(t') - \mathcal{C}^TF(t')^T B(t')\mathcal{C}F(t') ]+\
  \widetilde{ \mathcal{L}}{in}(x'',x'),
   \eea where $
\mathcal{L}in(x'',x')$ means that this expression is linear in
$x''$ and $x'$. Since we want to find $\bar{G} \in
\{\bar{A},\bar{B},\bar{C}\}$, and e.g. \bea\label{elim}
\bar{G}_{kl} = \bar{G}_{kl}(t'',t') = \frac{\partial^2
\bar{S}_0}{\partial x''_k
  \partial x''_l} = \frac12\;\frac{\partial^2
  \Big( \left[ x^T A\,\dot{x} + x^T B\, x \right]\mid_{t'}^{t''}\Big)}{\partial x''_k
  \partial x''_l} \,,  \eea
it is necessary to have the following properties.

\bigskip

\noi {\bf Lemma 3.4} {\it The following relations hold \bea && \nn
\frac{\partial \mathcal{C}^T}{\partial x''_k}
=\frac{1}{\tr}\,[\,(-1)^{k+1}\tr_{k,1},(-1)^{k+2}\tr_{k,2}, \dots,
(-1)^{k+2n}\tr_{k,2n}\,]\,, \\[3pt] && \nn
\frac{\partial \mathcal{C}^T}{\partial x'_k}
=\frac{1}{\tr}\,[\,(-1)^{n+k+1}\tr_{n+k,1},(-1)^{n+k+2}\tr_{n+k,2},
\dots, (-1)^{n+k+2n}\tr_{n+k,2n}\,]\,,\hp{20mm}\\[3pt] \nn &&
\frac{\partial (\mathcal{C}^T\,F^T(t''))}{\partial x''_k}=
\frac{\partial \mathcal{C}^T}{\partial x''_k}\,\,F^T(t'')
=[\,0,\dots,0,\stackrel{k}{1},0,\dots,0\,]\,,\\[3pt] \nn &&
\frac{\partial (C^T\,F^T(t'))}{\partial x''_k}= \frac{\partial
C^T}{\partial x''_k}\,\,F^T(t')=0=\frac{\partial C^T}{\partial
x'_k}\,\,F^T(t'')=\frac{\partial (C^T\,F^T(t''))}{\partial
x'_k}\;,\\[3pt] \label{sdif} && \frac{\partial (\dot{F}'C)}{\partial
x''_k}= \dot{F}'\frac{\partial \mathcal{C}}{\partial
x''_k}\,=\frac{1}{\tr}\,\,[\,\dot{\tr}_{k}(f'_1),\dot{\tr}_{k}(f'_2),
\dots, \dot{\tr}_{k}(f'_n)\,]\,, \\[3pt] \nn && \frac{\partial (\dot{F}''C)}{\partial
x''_k}= \dot{F}''\frac{\partial \mathcal{C}}{\partial
x''_k}\,=\frac{1}{\tr}\,\,[\,\dot{\tr}_{k}(f''_1),\dot{\tr}_{k}(f''_2),
\dots, \dot{\tr}_{k}(f''_n)\,]\,, \\[3pt] \nn &&
\frac{\partial (\dot{F}'C)}{\partial x'_k}= \dot{F}'\frac{\partial
\mathcal{C}}{\partial
x'_k}\,=\frac{1}{\tr}\,\,[\,\dot{\tr}_{n+k}(f'_1),\dot{\tr}_{n+k}(f'_2),
\dots, \dot{\tr}_{n+k}(f'_n)\,]\,, \\[3pt] \nn &&
\frac{\partial (\dot{F}''C)}{\partial x'_k}= \dot{F}''\frac{\partial
\mathcal{C}}{\partial
x'_k}\,=\frac{1}{\tr}\,\,[\,\dot{\tr}_{n+k}(f''_1),\dot{\tr}_{n+k}(f''_2),
\dots, \dot{\tr}_{n+k}(f''_n)\,]\,.\eea }

\bigskip
 Let us find now, the matrix elements $\bar{A}_{kl},\,
\bar{B}_{kl} $ and $ \bar{C}_{kl}$. Firstly we have \bea \nn
\frac{\partial^2 (x(t'')^T\,A''\,\dot{x}(t''))}{\partial x''_k
  \partial x''_l} &\jzg & \frac{\partial^2 (\mathcal{C}^TF''\,^T\,A''\,\dot{F}''\mathcal{C})}{\partial x''_k
  \partial x''_l}= \frac{\partial}{\partial x''_k} \left(\frac{\partial(\mathcal{C}^TF''\,^T\,A''\,\dot{F}''
  \mathcal{C})}{
  \partial x''_l}\right) \nn \\[5pt] \nn &\jzg& \frac{\partial}{\partial x''_k}
  \left(\frac{\partial\mathcal{C}^T}{\partial x''_l}\;
  F''\,^T\,A''\,\dot{F}''\mathcal{C}+\mathcal{C}^T
  F''\,^T\,A''\,\dot{F}''\frac{\partial\mathcal{C}}{\partial x''_l}\right)\\[5pt] \nn
 \mbox{$\mathcal{C}$ is linear in $x_k''$
  and $x_l''$} &\jzg& \frac{\partial\mathcal{C}^T}{\partial x''_l}\;
  F''\,^T\,A''\,\dot{F}''\frac{\partial\mathcal{C}}{\partial
  x''_k}+\frac{\partial\mathcal{C}^T}{\partial x''_k}
  F''\,^T\,A''\,\dot{F}''\frac{\partial\mathcal{C}}{\partial x''_l}\stackrel{(\ref{sdif})}{=}\\[3pt]
   &\jzg& \frac{1}{\tr}\;[\,0,\dots,0,\stackrel{l}{1},0,\dots,0\,]
  \,A''\,[\,\dot{\tr}_{k}(f''_1),\dot{\tr}_{k}(f''_2),
\dots, \dot{\tr}_{k}(f''_n)\,]^T\nn
\\[3pt] \nn && +\, \frac{1}{\tr}\;[\,0,\dots,0,\stackrel{k}{1},0,\dots,0\,]
  \,A''\,[\,\dot{\tr}_{l}(f''_1),\dot{\tr}_{l}(f''_2),
\dots, \dot{\tr}_{l}(f''_n)\,]^T\\[3pt]
   &\jzg& \frac{1}{\tr}\sum_t \Big(\alpha''_{lt}\dot{\tr}_k(f''_t)+\alpha''_{kt}\dot{\tr}_l(f''_t)\Big)\,.\eea

\bea \nn \frac{\partial^2 (x(t')^T\,A'\,\dot{x}(t'))}{\partial x''_k
  \partial x''_l} &\jzg & \frac{\partial^2 (\mathcal{C}^TF'\,^T\,A'\,\dot{F}'\mathcal{C})}{\partial x''_k
  \partial x''_l}= \frac{\partial}{\partial x''_k} \left(\frac{\partial(\mathcal{C}^TF'\,^T\,A'\,\dot{F}'
  \mathcal{C})}{
  \partial x''_l}\right) \\[5pt]
  &\jzg& \frac{\partial\mathcal{C}^T}{\partial x''_l}\;
  F'\,^T\,A'\,\dot{F}'\frac{\partial\mathcal{C}}{\partial
  x''_k}+\frac{\partial\mathcal{C}^T}{\partial x''_k}
  F'\,^T\,A'\,\dot{F}'\frac{\partial\mathcal{C}}{\partial x''_l}\stackrel{(\ref{sdif})}{=}
 0\,.\eea

\bea \nn \frac{\partial^2 (x(t')^T\,A'\,\dot{x}(t'))}{\partial x'_k
  \partial x'_l} &\jzg & \frac{\partial^2 (\mathcal{C}^TF'\,^T\,A'\,\dot{F}'\mathcal{C})}{\partial x'_k
  \partial x'_l}= \frac{\partial}{\partial x'_k} \left(\frac{\partial(\mathcal{C}^TF'\,^T\,A'\,\dot{F}'
  \mathcal{C})}{
  \partial x'_l}\right) \nn \\[5pt]
   &\jzg& \frac{1}{\tr}\sum_t \Big(\alpha'_{lt}\dot{\tr}_{n+k}(f'_t)+\alpha'_{kt}\dot{\tr}_{n+l}(f'_t)\Big)\,.\eea

\bea \nn \frac{\partial^2 (x(t'')^T\,A''\,\dot{x}(t''))}{\partial
x'_k
  \partial x'_l} &\jzg & \frac{\partial^2 (\mathcal{C}^TF''\,^T\,A''\,\dot{F}''\mathcal{C})}{\partial x'_k
  \partial x'_l}= \frac{\partial}{\partial x'_k} \left(\frac{\partial(\mathcal{C}^TF''\,^T\,A''\,\dot{F}''
  \mathcal{C})}{
  \partial x'_l}\right) \nn \\[5pt]   &\jzg& \frac{\partial\mathcal{C}^T}{\partial x'_l}\;
  F''\,^T\,A''\,\dot{F}''\frac{\partial\mathcal{C}}{\partial
  x'_k}+\frac{\partial\mathcal{C}^T}{\partial x'_k}
  F''\,^T\,A''\,\dot{F}''\frac{\partial\mathcal{C}}{\partial x''_l}\stackrel{(\ref{sdif})}{=}0\,.\eea

\bea \nn \frac{\partial^2 (x(t'')^T\,B''\,{x}(t''))}{\partial x''_k
  \partial x''_l} &\jzg & \frac{\partial^2 (\mathcal{C}^TF''\,^T\,B''\,{F}''\mathcal{C})}{\partial x''_k
  \partial x''_l}= \frac{\partial}{\partial x''_k} \left(\frac{\partial(\mathcal{C}^TF''\,^T\,B''\,{F}''
  \mathcal{C})}{
  \partial x''_l}\right) \nn \\[5pt] \nn
  &\jzg& \frac{\partial\mathcal{C}^T}{\partial x''_l}\;
  F''\,^T\,B''\,{F}''\frac{\partial\mathcal{C}}{\partial
  x''_k}+\frac{\partial\mathcal{C}^T}{\partial x''_k}
  F''\,^T\,B''\,{F}''\frac{\partial\mathcal{C}}{\partial x''_l}\stackrel{(\ref{sdif})}{=}\,\nn
\\[3pt] \nn &\jzg&  [\,0,\dots,0,\stackrel{k}{1},0,\dots,0\,]
  \,B''\,[\,0,\dots,0,\stackrel{l}{1},0,\dots,0\,]^T\nn
\\[3pt]  && \hp{-15mm} +\  [\,0,\dots,0,\stackrel{l}{1},0,\dots,0\,]
  \,B''\,[\,0,\dots,0,\stackrel{k}{1},0,\dots,0\,]^T=\beta''_{lk}+\beta''_{kl}\,, \\[3pt]
  \frac{\partial^2 (x(t')^T\,B'\,{x}(t'))}{\partial x'_k
  \partial x'_l} &\jzg & \beta'_{lk}+\beta'_{kl}\,,\\[3pt]
  \frac{\partial^2 (x(t'')^T\,B''\,{x}(t''))}{\partial x'_k
  \partial x'_l} &\jzg & \frac{\partial^2 (x(t')^T\,B'\,{x}(t'))}{\partial x''_k
  \partial x''_l}=0\,. \eea

\bea \nn \frac{\partial^2 (x(t'')^T\,A''\,\dot{x}(t''))}{\partial
x''_k
  \partial x'_l} &\jzg & \frac{\partial^2 (\mathcal{C}^TF''\,^T\,A''\,\dot{F}''\mathcal{C})}{\partial x''_k
  \partial x'_l}= \frac{\partial}{\partial x''_k} \left(\frac{\partial(\mathcal{C}^TF''\,^T\,A''\,\dot{F}''
  \mathcal{C})}{
  \partial x'_l}\right) \nn \\[5pt] \nn
  &\jzg& \frac{\partial\mathcal{C}^T}{\partial x'_l}\;
  F''\,^T\,A''\,\dot{F}''\frac{\partial\mathcal{C}}{\partial
  x''_k}+\frac{\partial\mathcal{C}^T}{\partial x''_k}
  F''\,^T\,A''\,\dot{F}''\frac{\partial\mathcal{C}}{\partial x'_l}\stackrel{(\ref{sdif})}{=}\,\nn
\\[3pt] \nn &\jzg&  \frac{1}{\tr}\;[\,0,\dots,0,\stackrel{k}{1},0,\dots,0\,]
  \,A''\,[\,\dot{\tr}_{n+l}(f''_1),\dot{\tr}_{n+l}(f''_2),
\dots, \dot{\tr}_{n+l}(f''_n)\,]^T\\[3pt]
   &\jzg& \frac{1}{\tr}\sum_t \alpha''_{kt}\dot{\tr}_{n+l}(f''_t)\,,
 \\[3pt] \frac{\partial^2 (x(t')^T\,A'\,\dot{x}(t'))}{\partial
x''_k
  \partial x'_l} &\jzg & \frac{\partial^2 (\mathcal{C}^TF'\,^T\,A'\,\dot{F}'\mathcal{C})}{\partial x''_k
  \partial x'_l}= \frac{\partial}{\partial x''_k} \left(\frac{\partial(\mathcal{C}^TF'\,^T\,A'\,\dot{F}'
  \mathcal{C})}{
  \partial x'_l}\right) \nn \\[5pt] \nn
  &\jzg& \frac{\partial\mathcal{C}^T}{\partial x'_l}\;
  F'\,^T\,A'\,\dot{F}'\frac{\partial\mathcal{C}}{\partial
  x''_k}+\frac{\partial\mathcal{C}^T}{\partial x''_k}
  F'\,^T\,A'\,\dot{F}'\frac{\partial\mathcal{C}}{\partial x'_l}\stackrel{(\ref{sdif})}{=}\,\nn
\\[3pt] \nn &\jzg&  \frac{1}{\tr}\;[\,0,\dots,0,\stackrel{l}{1},0,\dots,0\,]
  \,A'\,[\,\dot{\tr}_{k}(f'_1),\dot{\tr}_{k}(f'_2),
\dots, \dot{\tr}_{k}(f'_n)\,]^T\\[3pt]
   &\jzg& \frac{1}{\tr}\sum_t \alpha'_{lt}\dot{\tr}_{k}(f'_t)\,.\\[3pt]
   \frac{\partial^2 (x(t'')^T\,B''\,{x}(t''))}{\partial
x''_k
  \partial x'_l} &\jzg & \frac{\partial^2 (\mathcal{C}^TF''\,^T\,B''\,{F}''\mathcal{C})}{\partial x''_k
  \partial x'_l}= \frac{\partial}{\partial x''_k} \left(\frac{\partial(\mathcal{C}^TF''\,^T\,B''\,{F}''
  \mathcal{C})}{
  \partial x'_l}\right) \nn \\[5pt]   &\jzg& \frac{\partial\mathcal{C}^T}{\partial x'_l}\;
  F''\,^T\,B''\,{F}''\frac{\partial\mathcal{C}}{\partial
  x''_k}+\frac{\partial\mathcal{C}^T}{\partial x''_k}
  F''\,^T\,B''\,{F}''\frac{\partial\mathcal{C}}{\partial x'_l}\stackrel{(\ref{sdif})}{=}0\,,
\\[3pt] \frac{\partial^2 (x(t')^T\,B'\,{x}(t'))}{\partial
x''_k
  \partial x'_l} &\jzg & \frac{\partial^2 (\mathcal{C}^TF'\,^T\,B'\,{F}'\mathcal{C})}{\partial x''_k
  \partial x'_l}= \frac{\partial}{\partial x''_k} \left(\frac{\partial(\mathcal{C}^TF'\,^T\,B'\,{F}'
  \mathcal{C})}{
  \partial x'_l}\right) \nn \\[5pt]
  &\jzg& \frac{\partial\mathcal{C}^T}{\partial x'_l}\;
  F'\,^T\,B'\,{F}'\frac{\partial\mathcal{C}}{\partial
  x''_k}+\frac{\partial\mathcal{C}^T}{\partial x''_k}
  F'\,^T\,B'\,{F}'\frac{\partial\mathcal{C}}{\partial x'_l}\stackrel{(\ref{sdif})}{=}0\,. \label{last} \eea
Now, using (\ref{elim}), as well as (\ref{sdif})-(\ref{last}) we
have
\bigskip
 \noindent{\bf Theorem 3.5.} {\it The related
coefficients are: \bea \bar{A}_{kl}=\bar{A}_{kl}(t'',t')  &\jzg&
\frac{1}{2\,\tr}\sum_{t=1}^n
\Big(\alpha''_{lt}\dot{\tr}_k(f''_t)+\alpha''_{kt}\dot{\tr}_l(f''_t)\Big)+
\frac{\beta''_{lk}+\beta''_{kl}}2\;,\hp{17mm}\label{P3.3.1}\\[3pt]
  \bar{B}_{kl}=\bar{B}_{kl}(t'',t') &\jzg&  \frac{1}{2\tr}\sum_{t=1}^n \left(\alpha''_{kt}\dot{\tr}_{n+l}(f''_t)
  -\alpha'_{lt}\dot{\tr}_{k}(f'_t)\right) \hp{17mm}\label{P3.3.2}\\[3pt]
  \bar{C}_{kl}=\bar{C}_{kl}(t'',t') &\jzg&  \frac{-1}{2\,\tr}\sum_{t=1}^n
\Big(\alpha'_{lt}\dot{\tr}_{n+k}(f'_t)+\alpha'_{kt}\dot{\tr}_{n+l}(f'_t)\Big)-
\frac{\beta'_{lk}+\beta'_{kl}}2\;.\hp{17mm}\label{P3.3.3}\eea }

\section{Path Integrals on Real and $p$-Adic Spaces}

According to Feynman's path integral approach, discussed in the
Introduction, to obtain the complete transition amplitude from
$(x',t')$ to $(x'',t'')$  one has to take sum of amplitudes over
all possible trajectories $q(t)$ which interpolate between points
$(x',t')$ and $(x'',t'')$.   Any quantum path may be regarded as a
deviation $y(t)$ with respect to the classical one $x(t)$, i.e.
$q(t) = x(t) + y(t)$, where $y'=y(t')=0$ and $y''=y(t'')=0$. The
corresponding Taylor expansion of the quadratic action functional
$S[q]$ around classical path $x(t)$  is \bea
   S[q] &\jzg & S[x+y] = S[x] + \delta S[x] +  \frac{1}{2!}\; \delta^2 S[x] \label{4.1}
    \\[5pt] \nn &\jzg & S[x] +
   \frac{1}{2}\int_{t'}^{t''}  \left( \dot{y}_k \, \frac{\partial}{\partial \dot{q}_k} +
   y_k\, \frac{\partial}{\partial q_k}   \right)^2 L(\dot{q},q,t)dt .
\eea
Since our Lagrangian is  a polynomial up to quadratic order in
$\dot{q}_k$ and $q_k$, the terms with higher derivatives in (\ref{4.1})
are equal zero. Note also that the classical path $x(t)$ gives an
extremum of the action and hence we take $\delta S[x] = 0$.
According to (\ref{1.4}) and (\ref{1.9}), for any $v = \infty,
2,3,\cdots$, we can write
\bea
   {\cal K}_v (x'',t'';x',t') = \int
   \chi_v \left( -\frac{1}{h}\, S[x+y] \right)  {\cal D}y ,
    \label{4.2}
\eea
where  we replaced ${\cal D}q $ by  ${\cal D}y$, since $x$ is a
fixed classical trajectory. Due to (\ref{4.1}), the expression (\ref{4.2})
gains the more explicit form
\bea \label{4.3}
   \hspace{-10mm}{\cal K}_v (x'',t'';x',t') &\jzg&  \chi_v \left( -\frac{1}{h} \,
   \bar{S}(x'',t'';x',t') \right)\\[4pt]
&& \hspace{1mm} \nn
 \times\int_{y'\to 0,t'}^{y''\to 0,t''}
   \chi_v \left(- \frac{1}{2h} \int_{t'}^{t''}  \left( \dot{y}_k
   \, \frac{\partial}{\partial\dot{q}_k} +
    y_k \, \frac{\partial}{\partial q_k} \right)^2 L(\dot{q},q,t)dt
   \right) {\cal D}y,
    \eea
where we used $ y''=y'=0 , \quad$ $S[x] = \bar{S}(x'',t'';x',t')$.

\bigskip

\noindent{\textbf{Proposition 4.1.}} {\it ${\cal K}_v(x'',t'';x',t')$  has
the form
\bea
    {\cal K}_v(x'',t'';x',t') = N_v(t'',t') \chi_v \left( -\frac{1}{h}
    \, \bar{S}(x'',t'';x',t')\right) ,   \label{4.4}
\eea
where $N_v(t'',t')$ does not depend on end points $x''$ and
$x'$.}

\bigskip

\noi{\it Proof. \,\,} It follows from (\ref{4.3}). $ \qquad \Box $

\bigskip
 To compute $N_v(t'',t')$, let us note that (\ref{4.3}) can be
rewritten as \bea
   {\cal K}_v (x'',t'';x',t') = \chi_v \left( -\frac{1}{h}
   \, \bar{S}(x'',t'';x',t') \right) K_v (0,t'';0,t'),     \label{4.5}
\eea
where $ K_v (0,t'';0,t') =  K_v (y'',t'';y',t')
|_{y''=y'=0}\; $ and
\bea
 K_v (y'',t'';y',t')  =
   \int_{y',t'}^{y'',t''}
   \chi_v\left(- \frac{1}{h} \int_{t'}^{t''}
   \left[ \frac{1}{2}\, \dot{y}_k \, \alpha_{kl} \, \dot{y}_l
   +  \dot{y}_k \, \beta_{kl} \, {y}_l
   + \frac{1}{2}\, {y}_k \, \gamma_{kl} \, {y}_l           \right]
   \right) {\cal D}y.                                    \label{4.6}
\eea
Note that coefficients $\alpha_{kl} , $ $\, \beta_{kl}$ and
$\gamma_{kl}$ are those of the initial Lagrangian (\ref{3.1}). According
to (\ref{4.4}) and (\ref{4.5}) one has
\bea
  N_v(t'',t') = K_v(y'',t'';y',t')|_{y''=y'=0},             \label{4.7}
\eea
where
\bea
  K_v(y'',t'';y',t') = N_v(t'',t') \chi_v \left( -\frac{1}{h}
   \left[  \frac{1}{2}\, y''_k\, \bar{A}_{kl}\, y''_l +  y''_k\, \bar{B}_{kl}\, y'_l +
 \frac{1}{2}\, y'_k\, \bar{C}_{kl}\, y'_l             \right] \right).    \label{4.8}
\eea
with  $\bar{A}_{kl},$ $\, \bar{B}_{kl}$ and $\bar{C}_{kl}$ given by equations
(\ref{P3.3.1})-(\ref{P3.3.3}).

 To find the corresponding expression  for $N_v(t'',t')$ we shall
employ conditions (\ref{1.6}) and (\ref{1.7}). The unitary
condition (\ref{1.7}) now reads: \bea \int_{{\mathbb Q}^n_v}
\bar{K}_v(y'',t'';y',t') K_v(y,t'';y',t')\, d^n y' = \prod_{k=1}^n
\delta_v(y''_k -y_k). \label{4.9} \eea

\bigskip

\noindent{\bf Proposition 4.2.} {\it  The absolute value  of $N_v
(t'',t')$ in (\ref{4.4}) is \bea
  |N_v(t'',t')|_\infty  =
\left\vert \frac{1}{h^n} \det\left[ \frac{\partial^2 }{\partial
x''_k
\partial x'_l}\bar{S}_0(x'',t'';x',t')\right]\right\vert_v^{\frac{1}{2}}.  \label{4.10}
\eea }

\bigskip

\noi{\it Proof. \,\,} Substituting $ K_v(y'',t'';y',t')$ from
(\ref{4.8}) to (\ref{4.9}), and taking into account that the time
$t''$ is the same in points $y''$ and $y$, one obtains \bea
 && \nn \hspace{-10mm} |N_v(t'',t')|_\infty^2  \chi_v \left[ \frac{1}{2h}
 \bar{A}_{kl}(t'',t') (y''_k y''_l -y_k y_l) \right ] \int_{{\mathbb Q}_v^n} \chi_v
 \left[ \frac{1}{h} (y''_k - y_k) \bar{B}_{kl}(t'',t') y'_l \right]
 d^n y'= \hspace{10mm}\\[3pt]
 &&\hspace{17mm} = \prod_{k=1}^n \delta_v(y''_k -y_k). \hspace{-17mm}\label{4.11}
\eea
Using (\ref{2.13}),  one has
\bea && \nn \hspace{-10mm}
 |N_v(t'',t')|_\infty^2  \chi_v \left[ \frac{1}{2h}
 \bar{A}_{kl}(t'',t') (y''_k y''_l - y_k y_l) \right ]
 \left|\det \left[\frac{1}{h} \bar{B}_{kl} (t'',t')\right] \right|_v^{-1}
 \prod_{k=1}^n \delta_v(y''_k -y_k)\hspace{10mm}\\[3pt]
&&\hspace{17mm} = \prod_{k=1}^n \delta_v(y''_k -y_k).
\hspace{-17mm}     \label{4.12} \eea

Performing integration in (\ref{4.12}) over variable $y_k$, it
follows \bea
  |N_v(t'',t')|_\infty  =
\left\vert    \det \left(\frac{1}{h}  \bar{B}_{kl} (t'',t')\right)
\right\vert_v^{\frac{1}{2}}.          \label{4.13} \eea

Since  $\bar{B}_{kl} (t'',t')$ is the same for
$\bar{S}(x'',t'';x',t')$ and $\bar{S}(y'',t'';y',t')$, according
to (\ref{3.10}) one obtains (\ref{4.10}).        $ \qquad \Box $

We have now \bea
   N_v(t'',t') =  \left\vert
   \det \left(\frac{1}{h}\frac{\partial^2\bar{S}_0(y'',t'';y',t')}{\partial y''_k\partial y'_l}\right)
   \right\vert_v^{\frac{1}{2}}
   \mathcal{A}_v(t'',t') ,                                         \label{4.14}
\eea where $\vert \mathcal{A}_v(t'',t')\vert_\infty =1$ and $
\mathcal{A}_v(t'',t')$ remains to be determined explicitly. To
this end, we use condition (\ref{1.6}), which has now the form
\bea
 \int_{{\mathbb Q}_v^n} K_v(y'',t'';y,t) K_v(y,t;y',t') \, d^n y = K_v(y'',t'';y',t').
                                                            \label{4.15}
\eea

Inserting (\ref{4.8}) into (\ref{4.15}), where $N_v (t'',t')$ has
the form (\ref{4.14}), we get the following  equation:
\begin{small}
\bea\nn &&
N_v(t'',t')\chi_v\Big(-\frac1h\big(\frac12\,y''^T\bar{A}(t'',t')y''+
y''^T\bar{B}(t'',t')y'+\frac12 y'^T\bar{C}(t'',t')y'\big)\Big) =
N_v(t'',t)N_v(t,t')
\\[2pt] && \times\,\int_{{\mathbb Q}_v^n} \chi_v\Big(-\frac1h \big(\frac12\,y''^T\bar{A}(t'',t)y''+
y''^T\bar{B}(t'',t)y+ y^T \frac12 \bar{C}(t'',t)y+ \frac12\,y^T\bar{A}(t,t')y+
y^T\bar{B}(t,t')y'\nn \\[2pt] \nn &&+\ y'^T \frac12 \bar{C}(t,t')y'\big)\Big)d^ny = N_v(t'',t)N_v(t,t')
\chi_v\Big(-\frac{1}{2h}\big(y''^T\bar{A}(t'',t)y''+
y'^T\bar{C}(t,t')y'\big)\Big) \\[2pt] && \times\,\int_{{\mathbb Q}_v^n} \chi_v\Big(y^T\Big(\frac{\bar{C}(t'',t)+\bar{A}(t,t')}{-2h}\Big)y+
\Big(\frac{y''^T\bar{B}(t'',t)+
y'^T\bar{B}^T(t,t')}{-h}\Big)y\Big)d^ny=\mbox{\{Prop. 2.4\}}\nn\\[2pt] \nn &\hp{-2mm}=\hp{-2mm}&  N_v(t'',t)N_v(t,t')
\chi_v\Big(-\frac{1}{2h}\big(y''^T\bar{A}(t'',t)y''+
y'^T\bar{C}(t,t')y'\big)\Big)\Lambda_v(\alpha_1,\alpha_2,\dots,\alpha_n)\,|\det(2H)|_v^{-\frac12}\\[2pt]
&& \times\, \chi_v\Big(-\frac{1}{4}\, \big(z^T H^{-1} z\big)\Big),
 \nn\eea
\end{small}
where $\al_1,\dots,\al_n$ are eigenvalues of the symmetric matrix
\bea\nn H=\frac{\bar{C}(t'',t)+\bar{A}(t,t')}{-2h}\quad
\mbox{and}\quad z=\frac{y''^T\bar{B}(t'',t)+
y'^T\bar{B}^T(t,t')}{-h} . \eea Taking  into account (\ref{4.14}),
we can expect the following relations
\begin{small}\bea
 && \left\vert \det\Big(\frac{1}{h}\frac{\partial^2}{\partial y''
\partial y}\bar{S}_0(y'',y)\Big)
  \right\vert_v^{\frac{1}{2}}    \left\vert \det\Big(\frac{1}{h}
\frac{\partial^2}{\partial y\partial y'}
 \bar{S}_0(y,y')\Big)\right\vert_v^{\frac{1}{2}}  \vert \det
 (2H)\vert_v^{-\frac{1}{2}}\nn \\[3pt] &&
 = \left\vert \det\Big(\frac{1}{h}  \frac{\partial^2}{\partial y'' \partial y'}
  \bar{S}_0(y'',y')\Big)\right\vert_v^{\frac{1}{2}}, \label{id1}\\[12pt]
 && \chi_v\Big(-\frac1h\big(\frac12\,y''^T\bar{A}(t'',t')y''+
y''^T\bar{B}(t'',t')y'+\frac12 y'^T\bar{C}(t'',t')y'\big)\Big)
\nn \\[3pt] &&  =\chi_v\Big(-\frac{1}{2h}\big(y''^T\bar{A}(t'',t)y''+
y'^T\bar{C}(t,t')y'\big)\Big)\,\chi_v\Big(-\frac{1}{4}\, \big(z^T H^{-1}
z\big)\Big), \label{id2} \hp{8mm} \eea \end{small} which implies the third one
\bea
   \mathcal{A}_v(t'',t) \mathcal{A}_v(t,t')\Lambda_v (\alpha_1, \alpha_2, \cdots, \alpha_n) = \mathcal{A}_v(t'',t')
   . \label{id3}
\eea Let us introduce the following notations \bea\label{u1} \ba{ll}
U=(U_{ij}), \qquad & \hspace{-25mm}
\dst{U_{ij}=\frac12\;\frac{\dot{\tr}_i(f_j)(t,t')}{\tr (t,t')}-\frac12\;\frac{\dot{\tr}_{n+i}(f_j)(t'',t)}{\tr (t'',t)}\,,}\vp{3mm}\\
 \mathcal{H}=\bar{A}(t,t')+\bar{C}(t'',t)= A(t)\times U=(w_{ij})\,,\qquad & w_{ij}=\alpha^i\cdot U_j+\alpha^j\cdot U_i\,, \ea \eea
where $\alpha^i$ is $i-$th column of matrix $A$, and $U_j$ is $j-$th
row of matrix $U.$ By the multi-linearity of determinant one
reads\bea\nn \det \mathcal{H} = \sum_{i_1<i_2\dots < i_k\atop
j_1<j_2\dots < j_{n-k} } \det
[\alpha^{i_1},\dots,\alpha^{i_k},U^{j_1},\dots,U^{j_{n-k}}]\,
\det[U^{i_1},\dots,U^{i_k},\alpha^{j_1},\dots,\alpha^{j_{n-k}}]\,.
\eea Then one can see that above determinant is equal to \bea\nn
\det \mathcal{H} &\jzg&  2^n\det
[\,\alpha^{1},\alpha^2,\dots,\alpha^{n}\,]\,
\det[\,U^{1},U^2,\dots,U^{n}\,]\,+\, \mathcal{S} \\[3pt]
&\jzg&  \det A\, \det 2\, U\,+\, \mathcal{S} .\label{id1:0} \eea
Using Euler-Lagrange equations following ideas in \cite{grosj2}
one can find that $\mathcal{S}=0.$

For $X=(x_{ij}),\, Y=(y_{ij}),\, Z=(z_{ij})\,$ and
$\,W=(w_{ij})\,$ arbitrary matrices $(n,2\,n)$  ($n\in
\mathbb{N}$), and let $Y_{kl}$ be obtained from $Y$ replacing
$k-$th row of $Y$ by the $l-$th row of $X\,(k,l=1,2\dots,n)$.
Define the $(n,n)$ matrix $T=(t_{kl})$ by \bea\nn t_{kl}=\det
\left[\ba{c} Y_{kl}\\Z \ea\right]\, \det \left[\ba{c} Y \\W
\ea\right]-\det \left[\ba{c} Y_{kl}\\ W \ea\right]\, \det
\left[\ba{c} Y \\Z \ea\right]\,.\eea Then the following identity
holds \bea\label{di} \det T= \left(\det \left[\ba{c} Y\\Z
\ea\right]\right)^{n-1}\, \left(\det \left[\ba{c} Y\\W
\ea\right]\right) ^{n-1}\, \det \left[\ba{c} Z \\W \ea\right]\,
\det \left[\ba{c} X \\Y \ea\right]\,.\eea Using identity
(\ref{di}) one can find the following relation \bea\label{id1:1}
\det
2\,U=(-1)^n\frac{\tr(t'',t')}{\tr(t'',t)\,\tr(t,t')}\,\det\left[\ba{c}
F(t)\\ \dot{F}(t)\ea\right],\eea which combined with the relations
(\ref{sols}) (with $\mathcal{D}=1$), and (\ref{id1:0}) implies
\bea\label{id1:2} \det
\mathcal{H}=(-1)^n\frac{\tr(t'',t')}{\tr(t'',t)\,\tr(t,t')}\,,
\eea and since $\mathcal{H}=-2\,h H$, we have \bea\label{id1:3}
\det 2H =\det (\frac{-1}{h}\;\mathcal{H})=\frac{(-1)^n}{h^n}\;\det
\mathcal{H} =\frac{1}{h^n}
\frac{\tr(t'',t')}{\tr(t'',t)\,\tr(t,t')} . \eea Now, it is clear
that to prove (\ref{id1}) is enough to show that \bea\label{i4}
\det \bar{B}(t'',t') =\;\frac{1}{\tr(t'',t')}\,, \eea where
$\bar{B}(t'',t')$ is the matrix  $(\bar{B}_{kl}(t'',t')).$

In the case $n=1$, it is shown (see \cite{brank3}) that (\ref{i4})
holds and relations (\ref{id1})-(\ref{id3}) are satisfied with
\bea \mathcal{A}_v (t'',t') = \lambda_v \left( -\frac{1}{2 h}
\frac{\partial^2}{\partial y'' \partial
y'}\bar{S}_0(y'',t'';y',t') \right). \label{n1} \eea

 In the case $n=2$, using expressions (\ref{P3.3.2}) after long
calculations,  we find \begin{small}\bea\label{i5}\nn  \det
\bar{B}(t'',t')
&=&\bar{B}_{11} \bar{B}_{22} - \bar{B}_{12} \bar{B}_{21}=\frac{\left(\det A(t'')
\det\left[\ba{c} F(t'')\\ \dot{F}(t'')\ea\right] + \det A(t') \det\left[\ba{c} F(t')\\
\dot{F}(t')\ea\right]\right)\tr(t'',t') }{4\,\tr(t'',t')^2}\\[7pt] && +\
\frac{\mathrm{Tr}((A(t')\otimes
A(t''))\,\tilde{\tr})}{4\,\tr(t'',t')^2} =
\frac{1}{2\,\tr(t'',t')} + \frac{\mathrm{Tr}((A(t')\otimes
A(t''))\,\tilde{\tr})}{4\,\tr(t'',t')^2} ,  \eea \end{small} where
$ \tilde{\tr}=\left[\ba{rrrr} \tr_{2,1}  \tr_{4,1} & \tr_{2,1}
\tr_{4,2} & -\tr_{1,1}  \tr_{4,1} & -\tr_{1,1} \tr_{4,2}\\
\tr_{2,2}  \tr_{4,1} & \tr_{2,2}  \tr_{4,2} & -\tr_{1,2}
\tr_{4,1} & -\tr_{1,2}  \tr_{4,2}\\ -\tr_{2,1} \tr_{3,1} & -
\tr_{2,1}  \tr_{3,2} & \tr_{1,1}  \tr_{3,1} & \tr_{1,1}
\tr_{3,2}\\ -\tr_{2,2}  \tr_{3,1} & - \tr_{2,2} \tr_{3,2} &
\tr_{1,2}  \tr_{3,1} & \tr_{1,2}  \tr_{3,2}\ea\right]$
\vspace{3mm} and where
$\tr_{i,j}=\dot{\tr}_i({f}_j'),i=1,2;\,j=1,2$ and
$\tr_{i,j}=\dot{\tr}_i({f}_j''),i=3,4;\,j=1,2.$

From the properties of the function $\lambda_v$, we have \bea
\lambda_v\Big(\frac1a\Big)=\lambda_v\Big(a^2\,\frac1a\Big)=\lambda_v(a),
\, \, \lambda_v\Big( \frac1x + \frac1y \Big)=
\lambda_v\Big(\frac{x+y}{x\,y}
\Big)=\lambda_v\Big(\frac{x\,y}{x+y}\Big)=\frac{\lambda_v(x)\lambda_v(y)}{\lambda_v(x+y)}\,.\eea
Let us introduce the following notations \bea\nn
x=\triangle(t'',t) =\;\frac1{\det \bar{B}(t'',t)}\,,\quad
y=\triangle(t,t')=\;\frac1{\det \bar{B}(t,t')}\,,  \eea then  one
can write (as in $n=1$ case) \bea\label{lmul}
\lambda_v(\triangle(t'',t)+\triangle(t,t'))=\lambda_v(\triangle(t'',t'))\,.\eea

%Using $\lambda_p(-x)=\lambda_p(x)^{-1}$, we have \bea\label{Amul}
%\lambda_p(-\det
%H)=\frac{\mathcal{A}_p(t'',t)\,\mathcal{A}_p(t,t')}{\mathcal{A}_p(t'',t')}\,.
%\eea
Note that now \bea \Lambda_v (\alpha_1, \alpha_2, \cdots,
\alpha_n) = \lambda_v^{n-1}(1)\, \lambda_v (\det H)\,
\prod_{i<j\leq n} (\alpha_i , \alpha_j)_v   =
\frac{\mathcal{A}_v(t'',t')}{\mathcal{A}_v(t'',t)\,
\mathcal{A}_v(t,t')}. \label{H-Lambda} \eea Generally, product of
the Hilbert symbols can be $+1$ or $-1$, and we will take here
that it is $+1$, i.e. $\Lambda_v (\alpha_1, \alpha_2, \cdots,
\alpha_n) = \lambda_v^{n-1}(1)\, \lambda_v (\det H).$ Then we want
to show that $\mathcal{A}_v(t'',t')$ has the form
$\mathcal{A}_v(t'',t')= \lambda_v^{1-n}(1)\,
\lambda_v(-\xi\,\triangle(t'',t'))= \lambda_v^{1-n}(1)\,
\lambda_v(-\xi/\triangle(t'',t')),$ where $\xi=\frac{1}{(2h)^n}$.
Since \bea\nn\lambda_v(\det
H)&\hp{-2mm}=\hp{-2mm}&\lambda_v\Big(\frac{\xi\triangle(t'',t')}{\xi\,\triangle(t'',t)\xi\triangle(t,t')}\Big)=
\lambda_v\Big(\frac{\xi\triangle(t'',t)\,\xi\triangle(t,t')}{\xi\triangle(t'',t')}\Big)
\\ \nn &\hp{-2mm}=\hp{-2mm}&
\frac{\lambda_v(\xi\triangle(t'',t)) \,
\lambda_v(\xi\triangle(t,t'))}{\lambda_v(\xi\triangle(t'',t)+
\xi\triangle(t,t'))} =\frac{\lambda_v(\xi
\triangle(t'',t))\lambda_v(\xi
\triangle(t,t'))}{\lambda_v(\xi\triangle(t'',t'))}   \\
&\hp{-2mm}=\hp{-2mm}& \frac{\lambda_v(\xi \det
\bar{B}(t'',t))\lambda_v(\xi \det \bar{B}(t,t'))}{\lambda_v(\xi
\det\bar{B}(t'',t'))} = \frac{\lambda_v(- \xi
\det\bar{B}(t'',t'))}{\lambda_v(-\xi\det\bar{B}(t'',t))
\lambda_v(-\xi \det \bar{B}(t,t'))}.\label{lp2}\eea
So, if we compare (\ref{H-Lambda}) and (\ref{lp2}) %and (\ref{Amul})
we see that we obtain one class of solutions for \bea\label{Asol}
A_v(t'',t')=  \lambda_v^{1-n}(1)\, \lambda_v\left(
\frac{-1}{(2h)^n}\,\det \bar{B}(t'',t')\right).\eea

 In virtue  of the above evaluation one can formulate the
following

\bigskip

\noindent{\bf Theorem 4.3.}{\it The $v$-adic kernel ${\cal
K}_v(x'',t'';x',t')$ of the unitary evolution operator, defined by
(\ref{1.1}) and evaluated as the Feynman path integral, for
quadratic Lagrangians (\ref{3.1}) (and consequently, for quadratic
classical actions (\ref{3.7})) has the form}
\begin{small}\bea\nn
{\cal K}_v(x'',t'';x',t') &=& \lambda_v^{1-n}(1)\, \lambda_v\left(
\frac{-1}{(2h)^n} \det  \left( \frac{\partial^2}{\partial x''_k
\partial x'_l}\bar{S}_0(x'',t'';x',t')\right) \right) \left\vert
\det \left( \frac{1}{h} \frac{\partial^2}
{\partial x''_k \partial x'_l}\bar{S}_0(x'',t'';x',t') \right) \right\vert_v^{\frac{1}{2}}\\[7pt]
&&\,\times \chi_v\left( -\frac{1}{h} \bar{S}(x'',t'';x',t')  \right)
        \label{4.28}      \eea\end{small}
{\it and satisfies the general properties
(\ref{1.6})-(\ref{1.7})}.

\bigskip

\noi {\it Proof. \,\,}  The formula (\ref{4.28}) is a result of
the above analytic evaluation, and one has to show that this
expression also satisfies explicitly (\ref{1.8}).
%In fact, it is already shown that
%(\ref{1.6}) and (\ref{1.7}) are satisfied for the reduced
%Lagrangian (\ref{3.17}), and in the analogous way the proof
%extends to the general case (\ref{3.1}). The property (\ref{1.8})
%follows from \bea
% \lim_{t'\to t''} \lambda_v \left( -\frac{1}{2h} \frac{\partial^2 \bar{S}}
% {\partial x'' \partial x'}  \right) \left\vert  \frac{1}{h} \frac{\partial^2
% \bar{S}}{\partial x'' \partial x'}  \right\vert_v^{\frac{1}{2}}   \chi_v \left[
%\frac{1}{2h} \frac{\partial^2 \bar{S}}{\partial x'' \partial x'}
%\left( x''^2 - 2x''x' + x'^2 \right)  \right] =\delta_v (x'' - x').
% \label{4.29}
%\eea
$ \qquad \Box $
\bigskip

Starting from (\ref{4.28}) and using definition (\ref{2.10}) for
$\lambda_\infty$-function one can rederive well-known result in
ordinary quantum mechanics: \bea {\cal K}_\infty(x'',t'';x',t') =
\sqrt{ \frac{1}{(i h)^n} \, \det\left( -
\frac{\partial^2}{\partial x''_k \partial
x'_l}\bar{S}_0(x'',t'';x',t') \right)} \exp\left( \frac{2\pi i}{h}
\bar{S}(x'',t'';x',t')  \right). \label{4.30} \eea

\section{Adelic Path Integral}

Adelic path integral can be introduced as a generalization of
ordinary and $p$-adic path integrals. As adelic analogue of
(\ref{1.3}) it is related to eigenfunctions in adelic quantum
mechanics in the form \bea
 \psi_{\mathcal{P},\alpha}(x'',t'') = \int_{\mathbb A}
 {\cal K}_{\mathbb A}(x'',t'';x',t') \psi_{{\mathcal P},\alpha}(x',t') dx',  \label{5.1}
\eea where  $\psi_{{\mathcal P},\alpha}(x,t)$ has the form
(\ref{2.20}). Since the equation (\ref{5.1}) must be valid for any
set ${\mathcal P}$ of primes $p$, and adelic eigenstate is an
infinite product of real and $p$-adic eigenfunctions, it is
natural to consider adelic probability amplitude in the following
form: \bea
 {\cal K}_{\mathbb A}(x'',t'';x',t') =
 {\cal K}_\infty(x_\infty'',t_\infty'';x_\infty',t_\infty')  \prod_p
 {\cal K}_p(x_p'',t_p'';x_p',t_p'),                                \label{5.2}
\eea where $ {\cal
K}_\infty(x_\infty'',t_\infty'';x_\infty',t_\infty')$  and $ {\cal
K}_p(x_p'',t_p'';x_p',t_p')$ are probability amplitudes in
ordinary and $p$-adic quantum mechanics, respectively.

From (\ref{5.2}),   we see that one can introduce adelic path
integral as an infinite product of ordinary and $p$-adic path
integrals for all primes $p$. We  consider adelic Feynman's path
integral as a path integral on  an adelic space. Now we can
rewrite (\ref{5.2}) in the form \bea
  {\cal K}_{\mathbb A}(x'',t'';x',t') = \int_{x',t'}^{x'',t''}
  \chi_{\mathbb A} \left( -\frac{1}{h} S_{\mathbb A}[q] \right)
  {\cal D}_{\mathbb A}q,                                        \label{5.3}
\eea where $\chi_{\mathbb A}(x)$ is adelic additive character,
$S_{\mathbb A}[q]$ and  $ {\cal D}_{\mathbb A}q $ are adelic
action and the Haar measure, respectively. For practical
considerations, we define adelic path integral in the form \bea
  {\cal K}_{\cal A}(x'',t'';x',t') = \prod_v
\int_{x_v',t_v'}^{x_v'',t_v''} \chi_v \left( -\frac{1}{h}
\int_{t_v'}^{t_v''} L (\dot{q}_v,q_v,t_v) dt_v  \right) {\cal
D}q_v . \label{5.4} \eea Adelic Lagrangian is the infinite
sequence \bea
 L_{\mathbb A}(\dot{q},q,t) = (L(\dot{q}_\infty,q_\infty,t_\infty),
L(\dot{q}_2,q_2,t_2), L(\dot{q}_3,q_3,t_3),\cdots,
 L(\dot{q}_p,q_p,t_p),\cdots),                               \label{5.5}
\eea where $|L(\dot{q}_p,q_p,t_p)|_p \leq 1$ for all primes $p$
but a finite set $\mathcal P$ of them. Consequently, an adelic
quadratic Lagrangian looks like (\ref{5.5}), where each element
$L(\dot{q}_v,q_v,t_v)$ has the same form (\ref{3.1}).

Taking into account results obtained in the previous sections, we
can write adelic path integral for $n$-dimensional quadratic
Lagrangians (and consequently, quadratic classical actions) as $$
{\cal K}_{\mathbb A}(x'',t'';x',t') = \prod_v  \lambda_v^{1-n} (1)
\, \, \lambda_v\left[ -\frac{1}{(2h)^n} \det \left(
\frac{\partial^2}{\partial x_{(v)k}''
\partial x_{(v)l}'}\bar{S}_0(x_v'',t_v'';x_v',t_v') \right)\right]
$$
\bea \times\left\vert \det \left( \frac{1}{h} \frac{\partial^2}
{\partial x_{(v)k}''
\partial x_{(v)l}'}\bar{S}_0(x_v'',t_v'';x_v',t_v')\right)
\right\vert_v^{\frac{1}{2}} \chi_v\left( -\frac{1}{h}
\bar{S}(x_v'',t_v'';x_v',t_v') \right).
                                                      \label{5.6}
\eea

Note that vacuum state $\Omega(|x_p|_p)$ transforms as \bea
 \Omega(|x_p''|_p) = \int_{{\mathbb{Q}}_p}{\cal K}_p(x_p'',t_p'';x_p',t_p')
 \Omega(|x_p'|_p) dx_p' =
 \int_{{\mathbb Z}_p}{\cal K}_p(x_p'',t_p'';x_p',t_p') dx_p'.     \label{5.7}
\eea As a consequence of (\ref{5.7}) one has \bea
 \int_{{\mathbb Z}_p}{\cal K}_p(x_p'',t_p'';x_p,t_p) {\cal K}_p(x_p,t_p;x_p',t_p')
dx_p  = {\cal K}_p(x_p'',t_p'';x_p',t_p'),     \label{5.8} \eea
which may be viewed as an additional condition on $p$-adic path
integrals in adelic quantum mechanics for all but a finite number
of primes $p$. Conditions (\ref{5.7}) and (\ref{5.8}) impose a
restriction on a dynamical system to be adelic. It is practically
a restriction on time $t_p$ to have consistent adelic time $t$.

\section{Concluding Remarks}

Evaluating path integrals simultaneously on real and $p$-adic
$n$-dimensinal spaces, in the previous sections we derived some
general expressions related to probability amplitudes ${\cal
K}(x'',t'';x',t')$ in ordinary, $p$-adic and adelic quantum
mechanics. It has been done for Lagrangians $L(\dot{q},q,t)$ which
are polynomials at most the second degree in  dynamical variables
$\dot{q}_k$ and $q_k$, where $k=1,2,\cdots, n$.

It is worth pointing out that the formalism of ordinary and
$p$-adic path integrals can be regarded as the same at different
levels of evaluation, and the obtained results have the same form.
In fact, this property of number field invariance has to be
natural for general mathematical methods in theoretical physics
and fundamental physical laws ({\it cf.} \cite{volov4}).

\section*{Acknowledgements}

The work  on this paper was partially supported by the Serbian
Ministry of Science and Technological Development under contract
No. 144032.

\end{document}